\documentstyle[11pt,newpasp,twoside]{article}
\def\edcomment#1{\iffalse\marginpar{\raggedright\sl#1\/}\else\relax\fi}
\reversemarginpar
\input{psfig}
\begin{document}
\title{A catalogue of AGN observed by the PDS experiment on board BeppoSAX}
 \author{Matteo Guainazzi}
\affil{XMM-Newton Science Operation Center, VILSPA, ESA, Apartado 50727, E-28080. Madrid, Spain}
\author{Tim Oosterbroek}
\affil{Astrophysics Division, ESTEC, ESA, Postbus 299, NL-2200 AG Noordwijk, The Netherlands}

\begin{abstract}
The PDS experiment onboard BeppoSAX is the most
sensitive instrument ever above 10~keV. Preliminary results on
the complete catalogue of AGN observed with it are
presented in this {\it paper}.
\end{abstract}

The Phoswitch Detector System (PDS; Frontera et al. 1997)
- one of the four Narrow Field Instruments on board the
Italian-Dutch satellite BeppoSAX (Boella et al. 1997) -  is an
array of four independent NaI(Tl)/CsI(Na) scintillation detectors.
With an hexagonal collimated field of view of 1$\fdg$3,
the PDS was designed to ensure moderate
resolution spectroscopy
($\Delta E/E \le 15\%$ at 60~keV)
in the energy range between 13 and 200~keV.
The accurate control of the instrumental and celestial background
(at the level of one tenth of mCrab) and the unprecedented
sensitivity are the most outstanding properties of this
instrument, whose performances are likely to remain unsurpassed
for many years to come.

In this {\it paper} we describe the state-of-the-art
complete catalogue of Active Galactic Nuclei (AGN) observed
by the PDS. Thanks to the last minute extension of the
operational life of BeppoSAX, one year beyond the originally foreseen
switch-off (April 2001), this is still a work in progress.

In this {\it paper} we consider all the
PDS observations of AGN, whose data were publicly available
on March 2001, and which yielded
a $\ge$3$\sigma$ PDS detection. For
the typical exposure time of a BeppoSAX observation (50~ks),
this level corresponds to a count rate of about $0.15$~s$^{-1}$,
or 1~mCrab.

\section{AGN spectroscopy above 10~keV}

The main goal of AGN observations above 10~keV is the
direct observation of the nuclear emission of
absorbed Seyferts\footnote{In this paper we
define as {\it nuclear} the region surrounding
the supermassive black hole within a radius of
the order of 1~pc, {\it i.e.}, the putative
inner size of the putative molecular torus
(see, {\it e.g.}, Guainazzi et al. 2000a)}.
Type 2 Seyferts suffer
significant extinction in the X-ray band, mainly
due to photoelectric
absorption by cold matter.
As long as the absorbing column density, $N_H$, is lower then
a few $10^{24}$~cm$^{-2}$, PDS spectra are
basically unaffected by absorption, and
the intrinsic nuclear emission of
different flavors of AGN can be directly compared\footnote{Recent
BeppoSAX studies (Risaliti et al. 1999)
suggest that a sizable fraction of Seyferts are
totally thick to Compton-scattering,
with $N_H > 10^{25}$~cm$^{-2}$.
In this case the nuclear transmitted
flux is strongly suppressed and even
the PDS is of very little help.}.

The general astrophysical context of this study is
the unification scenario for Seyfert galaxies
(Antonucci 1993). If
the observational differences between absorbed and
unabsorbed AGN are indeed mainly due to the orientation
of systems, that otherwise share a common
nuclear engine, hard X-ray observations should allow us
to observe the emission from the innermost region of
the active nucleus, overcoming any "contamination" from
the surrounding environment. Any Seyfert type-dependent
differences
in the hard X-ray properties would undermine one of
the basic assumption of the unification scenario.

In Figure 1 we show the distribution function
\begin{figure}
\begin{center}
\hbox{
\hspace{-0.5cm}
\psfig{figure=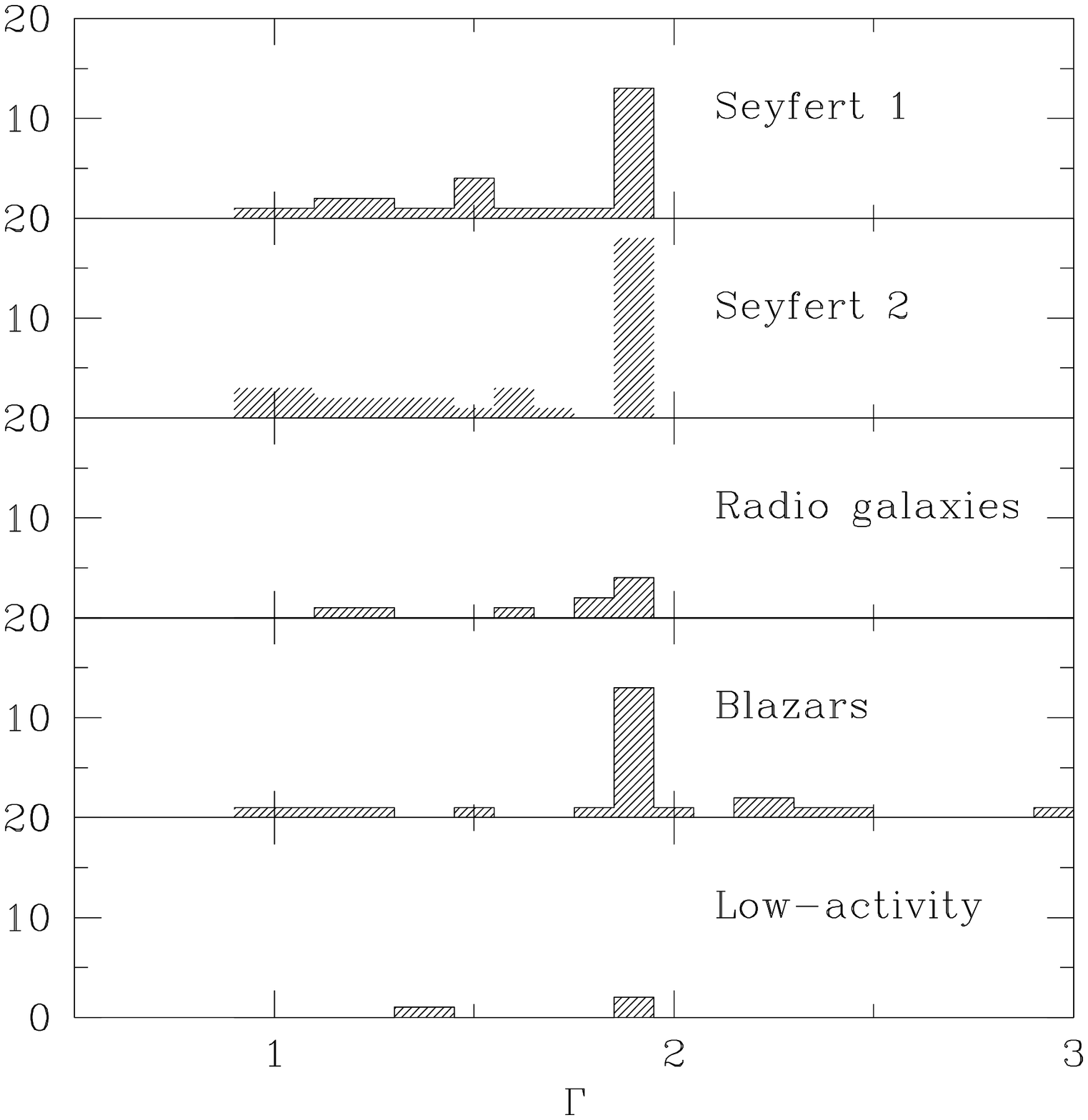,height=7.0cm,width=7.0cm}
\psfig{figure=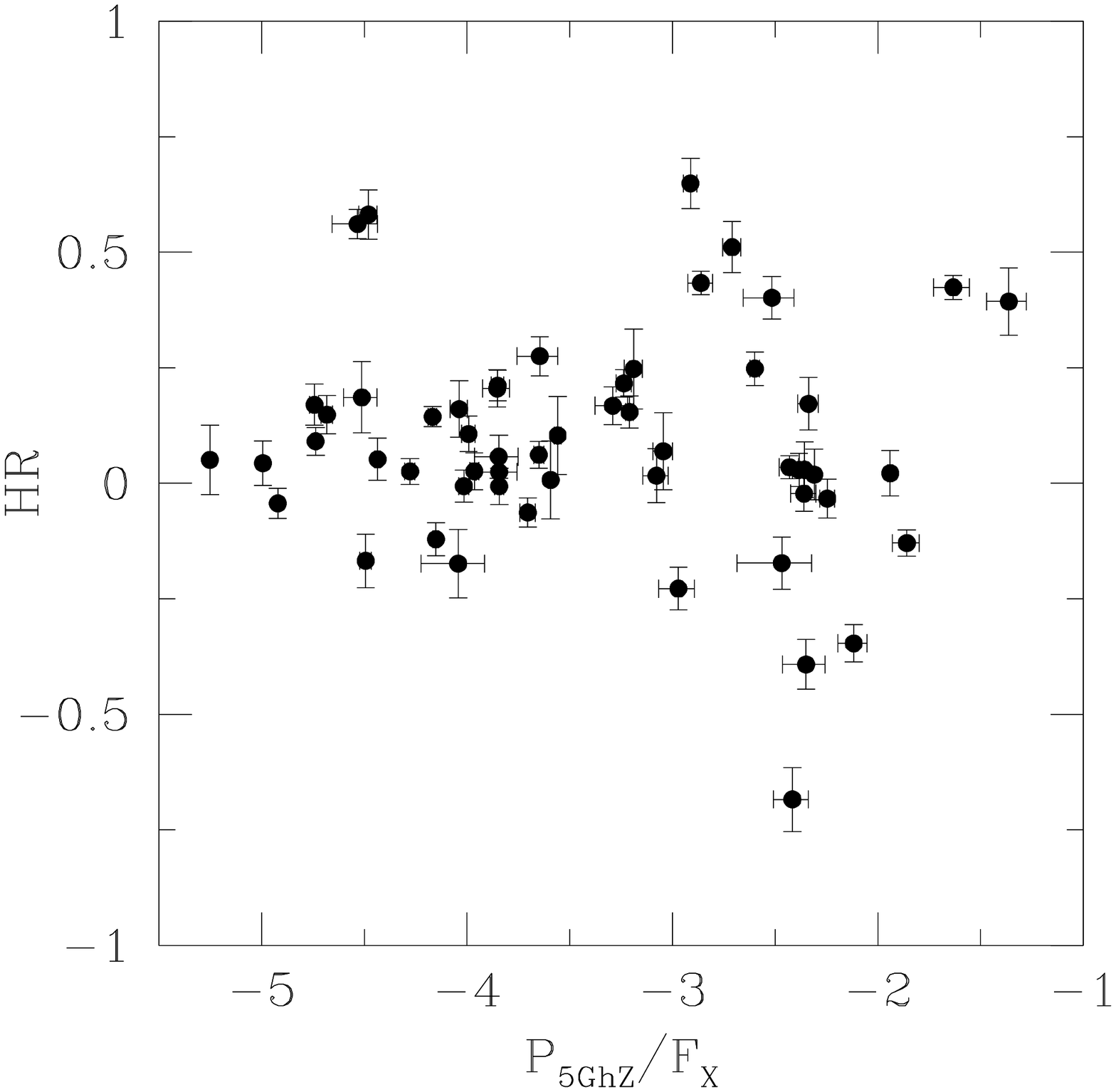,height=7.0cm,width=7.0cm}
}
\end{center}
\vspace{-1.0cm}
\caption{
{\it Left panel}: distribution function for the
PDS photon index of different classes of AGN.
{\it Right panel}: Hardness Ratio versus
ratio-to-X-ray power ratio for the whole
PDS AGN catalogue
}
\label{fig1}
\end{figure}
of the photon spectral indices for several classes
of AGN. PDS spectra have been fit with a simple
power-law model with high-energy cut-off (the
latter component being required
at a confidence level higher than 99\% only in 4 objects).
The mean of the distributions for type 1 and type 2
Seyferts are virtually indistinguishable:
$\langle \Gamma \rangle_{Sy1} = 1.678 \pm 0.014$
($\sigma_{Sy1} = 0.070$), and 
$\langle \Gamma \rangle_{Sy2} = 1.663 \pm 0.013$
($\sigma_{Sy2} = 0.069$), respectively.
For comparison, we show also the distribution
for a the sample of radio-loud quasars (radio galaxies and blazars):
$\langle \Gamma \rangle_{RG} = 1.75 \pm 0.04$
($\sigma_{RG} = 0.104$), and 
$\langle \Gamma \rangle_{Bl} = 1.90 \pm 0.03$
($\sigma_{Bl} = 0.128$), respectively.
The difference seems to be related to the skeweness of the
distribution around the maximum value.
In the {\it right panel} of Figure 1 we compare the PDS
hardness ratio [defined as $HR \equiv (H-S)/(H+S)$, where
$S$ and $H$ are the counts below and above 30~keV]
against the ratio between the 5~Ghz and the 20-200~keV fluxes, $R_{rx}$.
For $R_{rx} < 10^{-3}$, $HR$ is confined
\begin{figure}
\begin{center}
\hbox{
\hspace{-0.5cm}
\psfig{figure=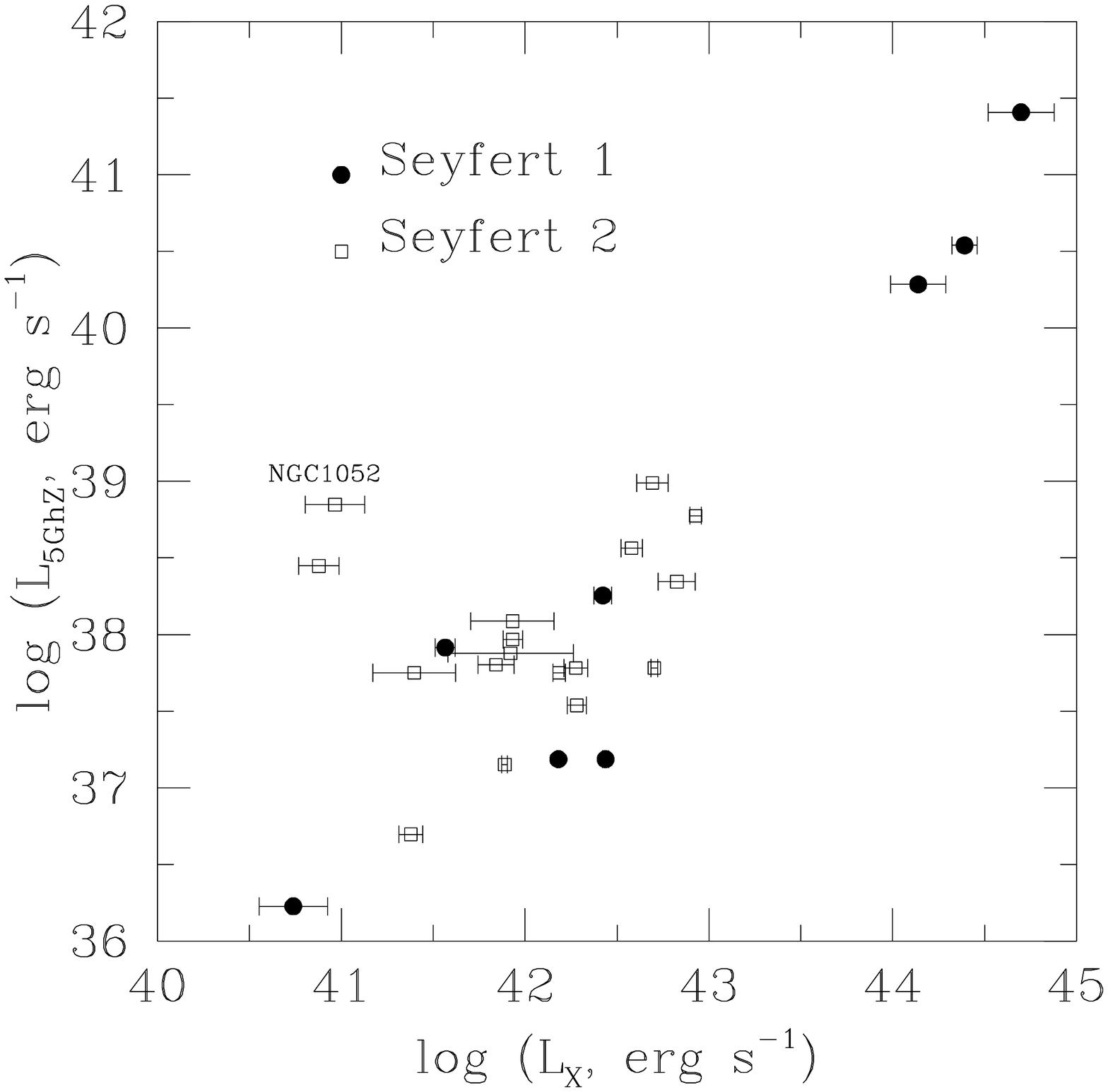,height=7.0cm,width=7.0cm}
\psfig{figure=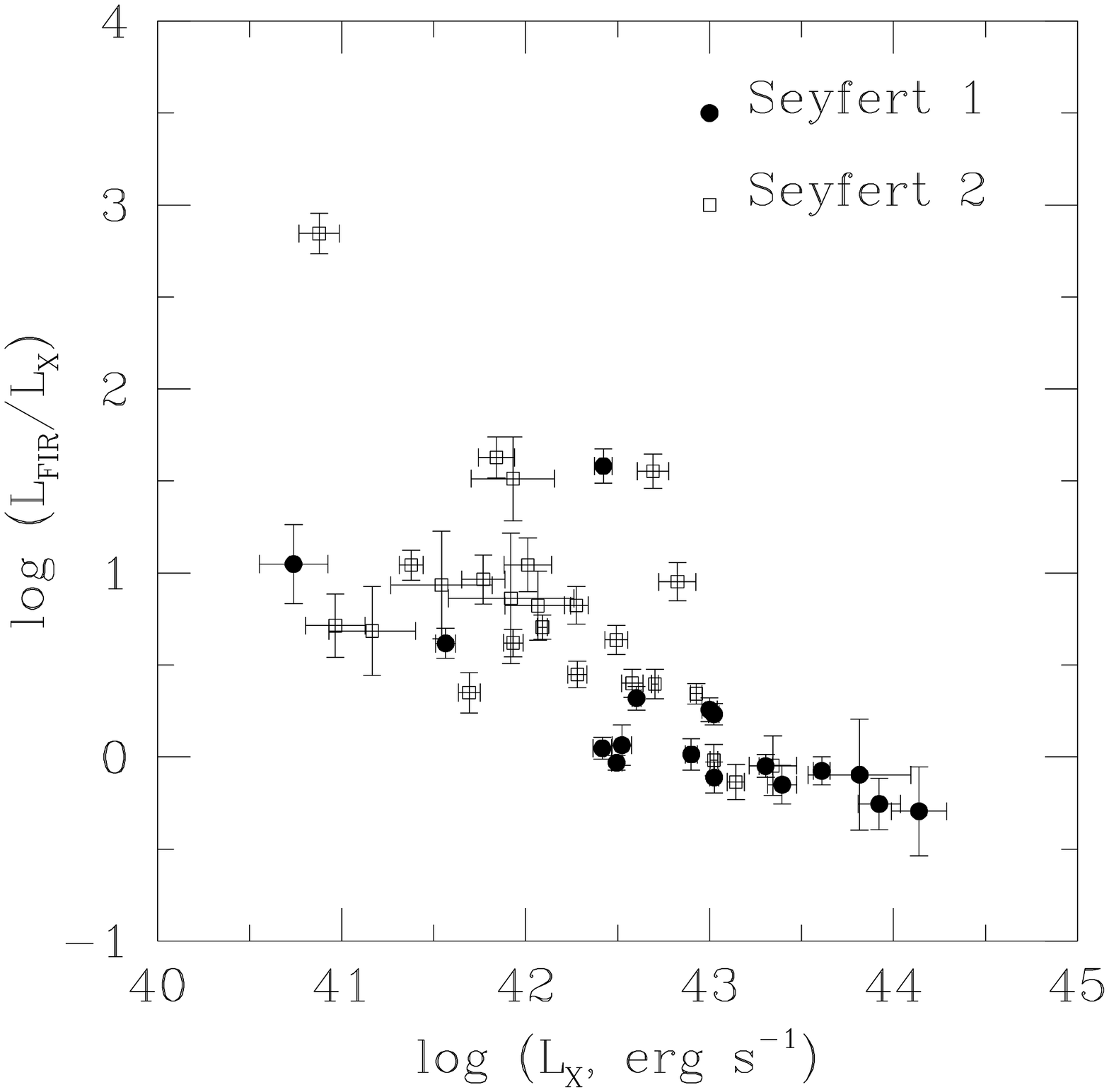,height=7.0cm,width=7.0cm}
}
\end{center}
\vspace{-1.0cm}
\caption{
{\it Left panel}: Radio versus X-ray luminosity for the
PDS AGN catalogue. The position of NGC~1052 is marked:
it is a LINER, whose X-ray emission is likely to be
dominated by an advection-dominated flow
(Guainazzi et al. 2000b)
{\it Right panel}: FIR-to-X-ray luminosity ratios
versus the X-ray luminosity for the same sample
}
\label{fig2}
\end{figure}
within a narrow band around $HR=0$,
whereas AGN with a stronger relative radio power
tend to suffer a substantially higher scatter of the
data points, likely to be due to the dichotomy of physical
processes (synchrotron and inverse Compton),
dominating the blazar emission in this energy
band.

\section{Multiwavelenght view}

The 20-200~keV luminosity, $L_X$ of the PDS Seyferts
shows clear correlations with
the radio (see the {\it left panel}
of Figure 2) and the FIR power
(Figure 3).
\begin{figure}[h]
\begin{center}
\hbox{
\psfig{figure=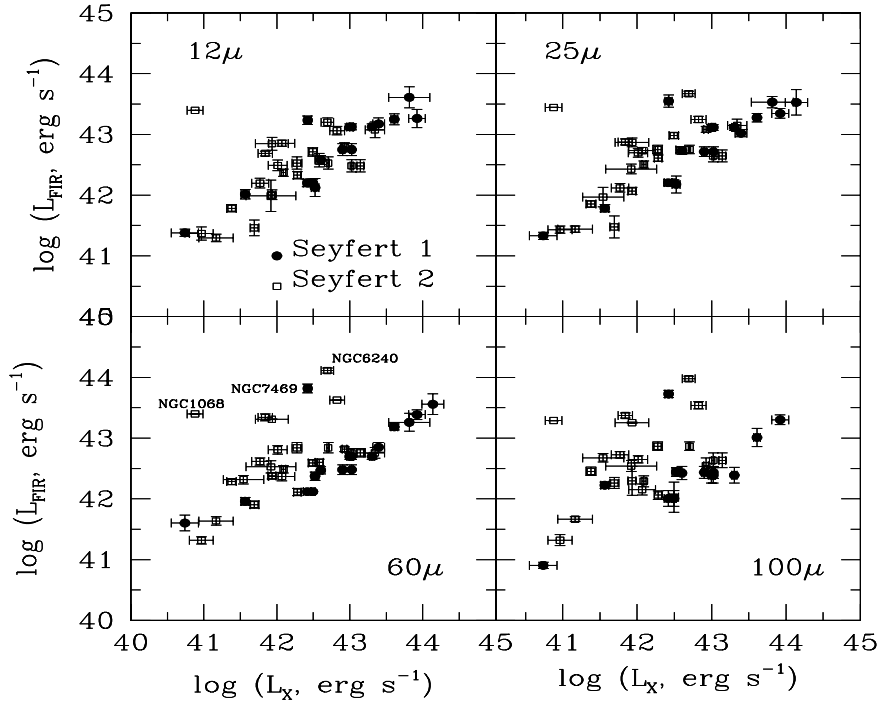,height=10.0cm,width=12.0cm}
}
\end{center}
\vspace{-1.5cm}
\caption{
FIR versus X-ray correlation for the Seyfert of the PDS catalogue
}
\label{fig3}
\end{figure}
The former correlation suggests a common nuclear origin
for both components. Indeed, at least 50\% of the 12\micron\
Seyfert sample show unresolved radio emission
on the sub-arcsecond scale (Thean et al. 2000).
The correlation between $L_X$ and $L_{FIR}$ is
at a significance level higher than 99.9\% for
all IRAS bands and for both type 1 and type 2
objects, except for the 100$\micron$ band
in Seyfert 1s ($\simeq 99\%$). This
supports the idea that the bulk of the FIR
emission in (bright) Seyferts is due to
reprocessing by optically-thick matter,
{\it e.g.} the molecular
torus surrounding the active nucleus in
the unification scenarios. Some objects exhibit, however,
an excess of FIR emission in comparison with the
members of their class of corresponding
X-ray luminosity. Three of these outliners
(NGC~1068, NGC~7469, and NGC~6240) are well
known to host intense nuclear starbursts, which
may well explain the IR excess emission in terms
of dust heating by young hot stars in regions
in intense star formation. Seyfert 1s and
Seyfert 2s do not seem to occupy the same
loci in the FIR vs. X-ray plane. As shown in
the {\it right panel} of Figure 2, for
$L_X < 10^{43}$~erg~s$^{-1}$ Seyfert~2s tend to have
on the average
a higher FIR-to-X-ray luminosity ratio than
Seyfert~1s, whereas this difference
vanishes for high-luminosity objects.
This would point to an increasingly
important starburst contribution to the
energy budget for decreasing AGN luminosity
and increasing AGN "obscureness".
Caution must be, however, used in the interpretation
of this result, because
the type 1 and 2 samples are not well matched in X-ray
luminosity.

\vspace{0.25cm}

\noindent
The authors acknowledge stimulating discussions with A.Zezas.

\end{document}